\documentclass{sig-alternate} 
\usepackage{mathptmx} 

\newcommand{\ignore}[1]{}
\usepackage{fancyhdr}
\usepackage[normalem]{ulem}
\usepackage[hyphens]{url}
\usepackage{microtype}
\usepackage{todonotes}
\usepackage{bbold}

\usepackage{scalerel}
\usepackage{tikz}
\usetikzlibrary{svg.path}
\definecolor{orcidlogocol}{HTML}{A6CE39}
\tikzset{
  orcidlogo/.pic={
    \fill[orcidlogocol] svg{M256,128c0,70.7-57.3,128-128,128C57.3,256,0,198.7,0,128C0,57.3,57.3,0,128,0C198.7,0,256,57.3,256,128z};
    \fill[white] svg{M86.3,186.2H70.9V79.1h15.4v48.4V186.2z}
                 svg{M108.9,79.1h41.6c39.6,0,57,28.3,57,53.6c0,27.5-21.5,53.6-56.8,53.6h-41.8V79.1z M124.3,172.4h24.5c34.9,0,42.9-26.5,42.9-39.7c0-21.5-13.7-39.7-43.7-39.7h-23.7V172.4z}
                 svg{M88.7,56.8c0,5.5-4.5,10.1-10.1,10.1c-5.6,0-10.1-4.6-10.1-10.1c0-5.6,4.5-10.1,10.1-10.1C84.2,46.7,88.7,51.3,88.7,56.8z};
  }
}
\newcommand\orcidicon[1]{\href{https://orcid.org/#1}{\mbox{\scalerel*{
\begin{tikzpicture}[yscale=-1,transform shape]
\pic{orcidlogo};
\end{tikzpicture}
}{|}}}}
\usepackage{hyperref} 

\title{Sidebar: Scratchpad Based Communication Between CPUs and Accelerators} 
\author {Ayoosh Bansal\orcidicon{0000-0002-4848-6850}\thanks{Equal Contribution},\hspace{0.3em} Chance Coats\footnotemark[1], \hspace{0.3em}Evan Lissoos\footnotemark[1] \hspace{0.3em}and Benjamin Schreiber\footnotemark[1]\\
University of Illinois at Urbana-Champaign, USA\\
\{ayooshb2, cccoats2, lissoos2, bjschre2\}@illinois.edu
}

\begin{document}
\maketitle
\pagestyle{plain}

\begin{abstract}
Hardware accelerators for neural networks have shown great promise for both performance and power. These accelerators are at their most efficient when optimized for a fixed functionality. But this inflexibility limits the longevity of the hardware itself as the underlying neural network algorithms and structures undergo improvements and changes. We propose and evaluate a flexible design paradigm for accelerators with a close coordination with host processors. The relatively static matrix operations are implemented in specialized accelerators while fast-evolving functions, such as activations, are computed on the host processor. This architecture is enabled by a low latency shared buffer we call \emph{Sidebar}. Sidebar memory is shared between the accelerator and host, exists outside of program address space and holds intermediate data only. We show that a generalised DMA dependent flexible accelerator design performs poorly in both perf and energy as compared to an equivalent fixed function accelerator. Sidebar based accelerator design achieves near identical performance and energy to equivalent fixed function accelerator while still providing all the flexibility of computing activations on the host processor.
\end{abstract}

\section{Introduction}

The rise in usage of deep neural networks has lead to unique computational demands on modern systems. Initially deployed on CPUs, neural networks have since moved to GPUs and FPGAs. Today, there are dedicated hardware blocks for neural networks in widely-available commodity hardware, including the latest SoCs developed by Apple, Qualcomm, and others~\cite{hwaccel}. Accelerators have also found their way into the data center, such as Google's TPU~\cite{tensorflow,tpu} and dominate GPUs in performance per watt for inference tasks.

Accelerators for deep learning excel at matrix multiplication, which forms the most computationally expensive portion of many modern models. However, these models also frequently involve non-linear activation functions. Some are comparatively easy to compute, such as ReLU, but others require special functions like tanh that are more expensive or require space for lookup tables. Many of these activation functions could be implemented in logic within the accelerator, but this approach lacks flexibility. While many advancements in deep learning will continue to map onto matrix operations, these non-linearities are more liable to change in the future and break hardware compatibility. Another option is to perform these operations on the CPU while keeping the matrix math on the accelerator. However, this requires costly DMA operations.

\begin{figure}[h]
\centering
\includegraphics[keepaspectratio,width=0.9\linewidth]{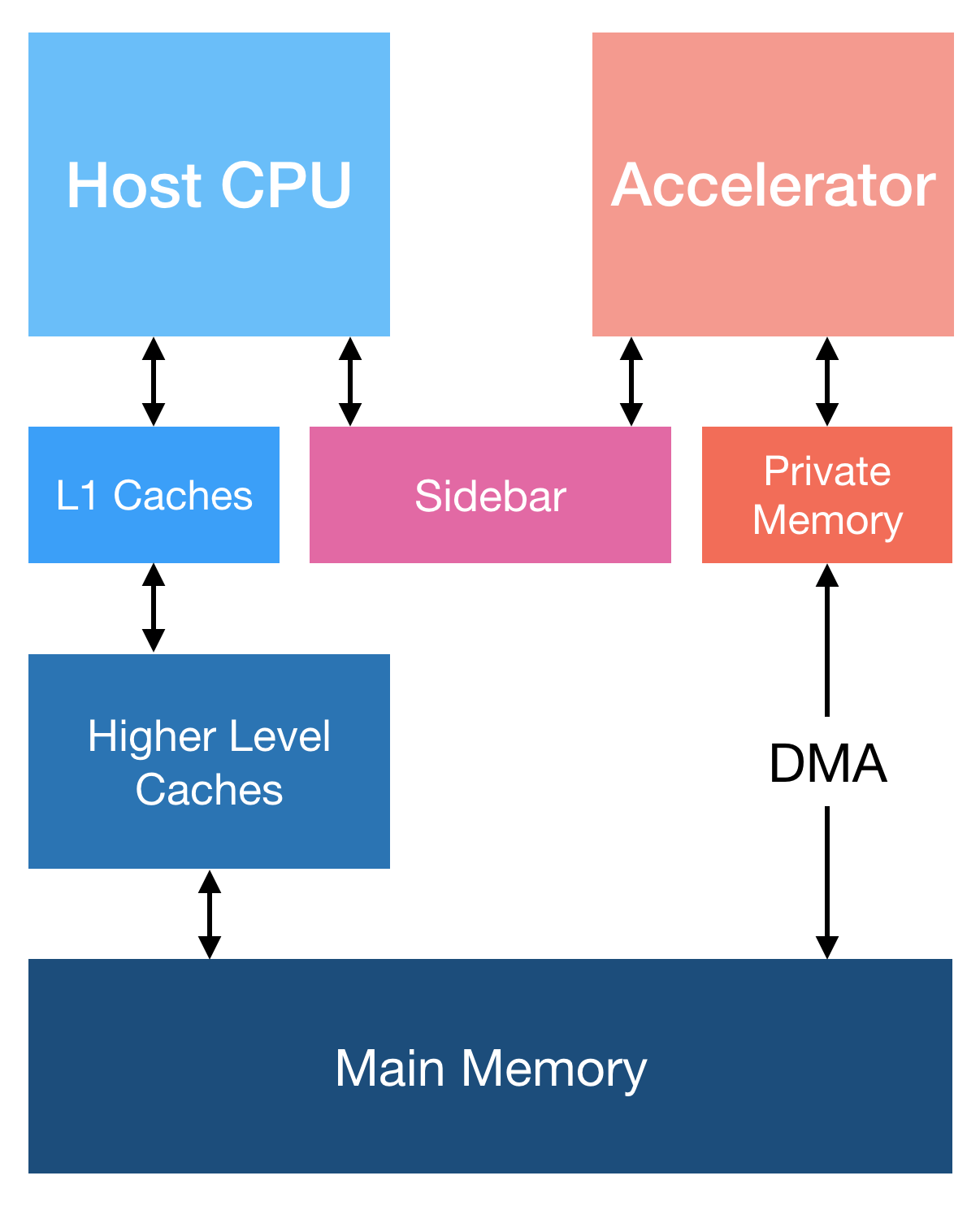}
\caption{\label{fig:sys}Proposed System Model}
\end{figure}

We evaluate the usage of a specialized buffer, called \emph{Sidebar} at the L1 level sitting between a CPU core and an accelerator block, see Figure~\ref{fig:sys}. The  accelerator will continue to have data pushed in through DMA operations to its private memory. Once loaded, the accelerator can perform computations on the data. If the accelerator is required to perform a computation that is expensive and not implemented in hardware, it can instead move this computation back to the host processor. The accelerator will copy the intermediate results it has computed into the \emph{Sidebar} and inform the host processor that it should perform some function on the \emph{Sidebar} data. The host processor performs the computation using all available execution resources, including vector units and complex arithmetic units. Then, the CPU sends the data back into the \emph{Sidebar} for the accelerator to use and continue execution.

Consider the example of accelerating a neural network. In our scheme, the CPU will initiate a neural network operation, such as a forward pass, on the accelerator. The accelerator will perform some matrix computations and at some point write intermediate values that require the application of the activation function into the \emph{Sidebar}. The CPU will compute the activation functions, write the results to the \emph{Sidebar} and then indicate to the accelerator that it may proceed. The accelerator can repeat this process until the neural network operation has completed.
\section{Background and Motivation}
\label{sec:background}
In this section we provide a brief introduction on neural networks plus the role and types of activation functions. 
\subsection{Neural Networks}\label{ssec:neuralnets}

Neural networks are a class of computational models. They have seen usage in a variety of domains, including image classification, translation, finance, autonomous vehicles, and more. In their most basic form, neural networks consist of compositions of linear predictors and activation functions. A linear predictor is a weighted average of its inputs plus a biasing term. The decision boundary is the sign of the output. Linear predictors themselves are used in some machine learning tasks, particularly regression tasks, but have limited representational power.

Compositions of linear predictors are themselves linear predictors. In order to increase the representational power of this composition, some non-linearities must be injected. The outputs of each linear predictor (each "fully connected layer") are passed into an activation function, whose result is then used as the input to another linear predictor ("layer"). In theory, a two layer network with a reasonable activation function can approximate any function arbitrarily well~\cite{Cybenko1989}. In practice, deeper networks are used for ease of training, as they require fewer total weights~\cite{telgarsky16}.

Modern neural networks frequently make use of more than just linear predictors. Convolutional layers perform a convolution on the input using a small kernel of weights. Pooling layers reduce the size of their input by replacing a sliding window over the input with a single entry in the output according to some algorithm, frequently either the max or the average. Other types, such as recurrent layers and dropout layers, have seen usage in some domains.

\subsection{Activations}\label{ssec:activations}

A wide variety of activation functions have been used, with their relative popularity changing over time. Early research in neural networks focused on perceptron networks and used the heaviside function. Later research focused on the sigmoid function, which endured for several decades. The hyborbolic tangent function was also used during this time. After rising to popularity with its usage in the winning entry of Imagenet 2012~\cite{alexnet}, the relu function remains the most popular activation function today. Many variants of it have been proposed and adopted to varying degrees, and more are sure to be developed in the future. Activation functions are distinct from other parts of a neural network in that they cannot be expressed as a matrix operation, and thus require special hardware. If new activation functions come into use, existing accelerators may not be able to implement them without hardware modification.

\begin{table}[htbp]
\large
\caption{Common Activation Functions}
\label{tab:activations}
\begin{center}
\bgroup
\def\arraystretch{2}
\begin{tabular}{l|r} 
\textbf{Name} & \textbf{Formula} \\
\hline
Heaviside     & $f(x) = \mathbb{1}[x > 0]$ \\
\hline
tanh          & $f(x) = \text{tanh}(x)$ \\
\hline
Sigmoid       & $f(x) = 1 / (1+e^{-x})$ \\
\hline
ReLU          & $f(x) = \text{max}(0,x)$ \\
\hline
Leaky ReLU    & $f(x) = \begin{cases}
                 x, & \text{if x > 0}\\
                 0.01x, & \text{otherwise}
                \end{cases}$ \\
\hline
ELU           & $f(x) = \begin{cases}
                 x, & \text{if x > 0}\\
                 a(e^x - 1), & \text{otherwise}
                \end{cases}$ \\
\hline
Softplus      & $f(x) = \log(1 + e^x)$

\end{tabular}
\egroup
\end{center}
\end{table}
\subsection{Motivation}

\begin{figure}[ht]
\centering
\includegraphics[width=\linewidth]{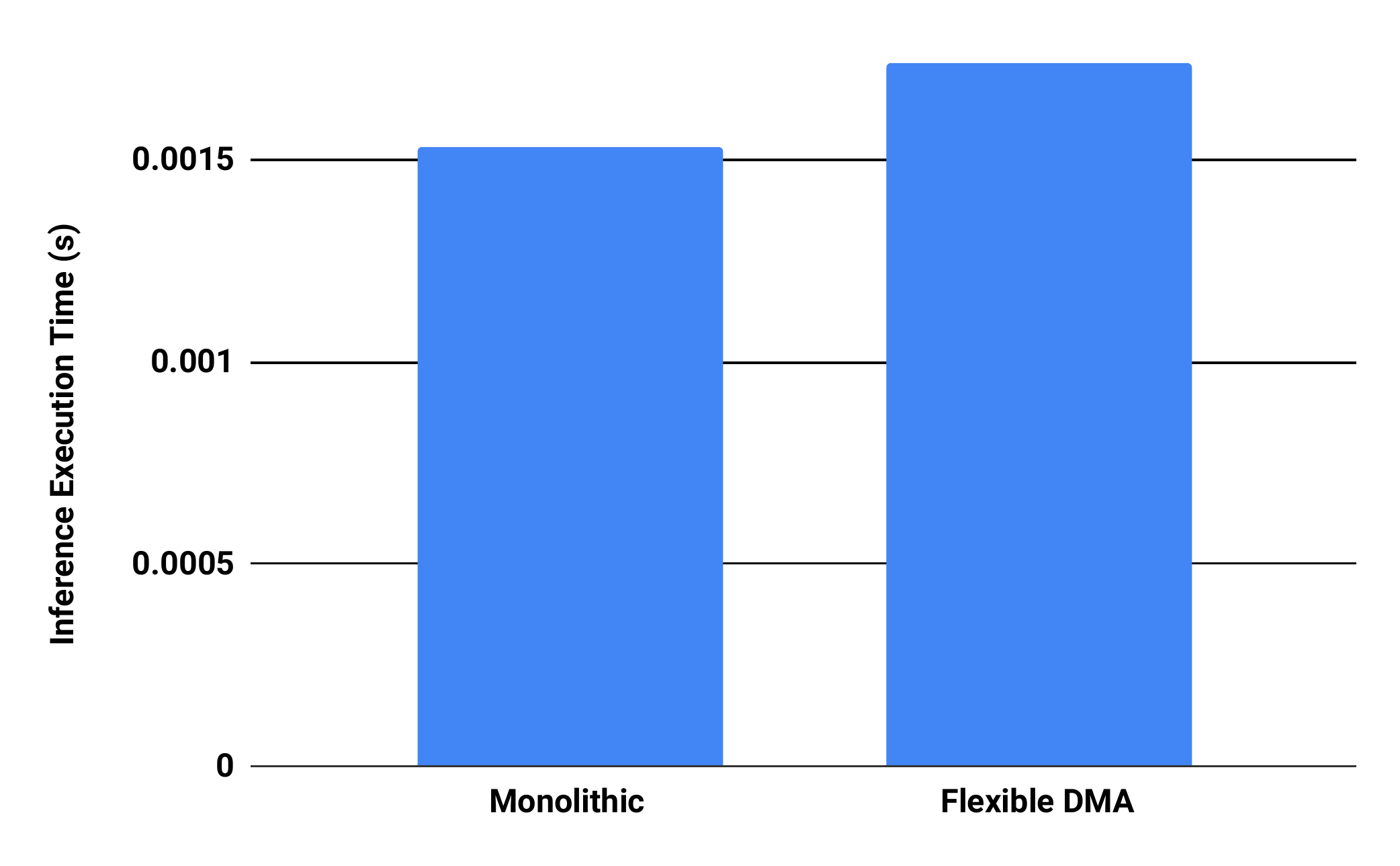}
\caption{Monolithic vs Flexible DMA Inference Performance}
\label{fig:mot_perf}
\end{figure}

\begin{figure}[ht]
\centering
\includegraphics[width=\linewidth]{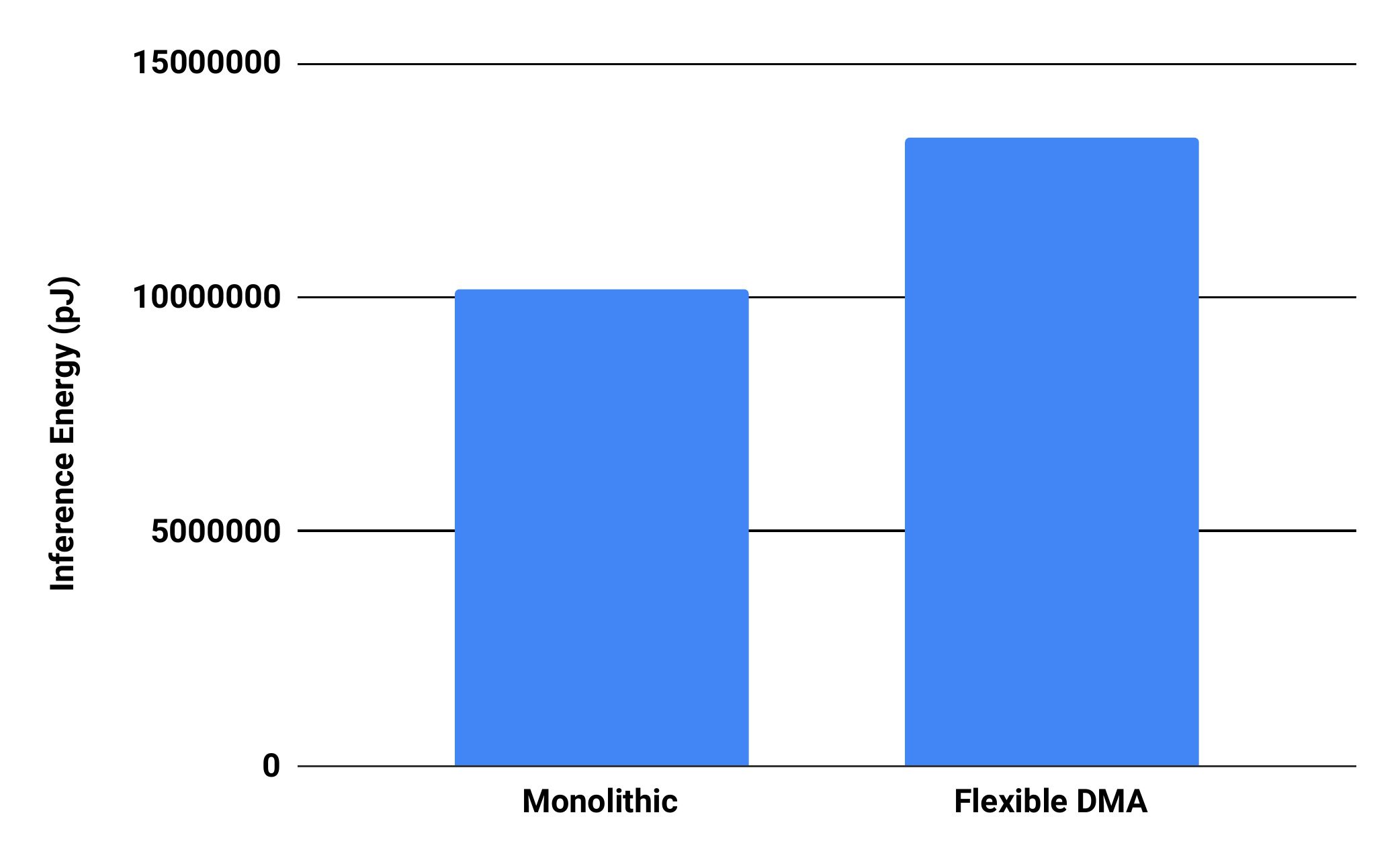}
\caption{Monolithic vs Flexible DMA Inference Energy}
\label{fig:mot_energy}
\end{figure}

Consider a large monolithic accelerator, like Figure~\ref{fig:accel}, which does includes a lot of layers and activation functions. As discussed before, the activation functions tend to change over time and with any small change in the algorithm of the accelerator, the complete hardware IP becomes obsolete and would need expensive engineering efforts to update. A better design is to have smaller primitives, S1-S5 in Figure~\ref{fig:accel}, which are relatively static and do the activation computations on a processor. Though, the interface for such a flexible system is a key problem. Several interfaces have been explored \cite{interfaceexploration,spandex} but in  most cases the data movement costs make this flexible design prohibitively expensive\cite{shao2015toward}. With the system described later in Section \ref{sec:impl}, we compare the monolithic accelerator and the flexible DMA accelerators performance and energy in Figures~\ref{fig:mot_perf} and~\ref{fig:mot_energy}. Clearly the naive Flexible design is prohibitively expensive and requires a better interface to move data to and from the host processor.

\section{System Considerations}
\label{sec:considerations}

The goal of this work is to provide a mechanism which can reduce the system overheads associated with fine-grained cooperation between an accelerator and a host CPU. We accomplish this by leveraging a tightly-coupled buffer, referred to as \emph{Sidebar}, as the point of contact between CPU and accelerator. Much like in a courtroom setting, the \emph{Sidebar} allows host processor and accelerators to have a quick communication invisible to the rest of the memory system. This mechanism enables the development of flexible accelerator hardware with better longevity than fixed function accelerators. The system model is portrayed in Figure~\ref{fig:sys}.

In order to meet this goal, careful considerations must be made. An explanation of how the \emph{Sidebar} will be accessed is described in Section \ref{ssec:DMA}, while its ability to enable fine-grained accelerator-CPU cooperation is detailed in Section \ref{ssec:coop}. Further, interactions between our work and existing processor coherence mechanisms are discussed in Section \ref{ssec:coherence}, while software integration with a host operating system is discussed in Section \ref{ssec:integration}.

\subsection{Accessing the Sidebar}\label{ssec:DMA}
The usefulness of the \emph{Sidebar} in our design relies upon the addition of at least two instructions to the host processor. These instructions, \textit{sbLD} and \textit{sbST}, allow the processor to load from and store to the Sidebar memory respectively. We use specialized instructions, instead of a memory mapping, to further isolate the \emph{Sidebar} from the main memory space and avoid coherence issues, discussed in \ref{ssec:coherence}.

The \textit{sbLD} instruction will primarily be used after the accelerator has completed an intermediate task and has signaled this completion to the host. The processor uses \textit{sbLD} to move intermediate data from the \emph{Sidebar} into its own register file, where it can then perform arbitrary computation on it. The \textit{sbST} instruction will be used to return data that the CPU has performed additional computation on to the \emph{Sidebar}.

In both cases, data placement is explicitly managed. There must be agreement between the accelerator and host code at compile-time on where data will be located within the \emph{Sidebar}, and how it will be arranged. This does place some additional demands on the programmer, but we believe this can be mitigated by simple compilation tools or frameworks, which we leave to future work.

The accelerator may access the \emph{Sidebar} in a similar manner. We do not allow the accelerator and the host processor to access the \emph{Sidebar} simultaneously, and we prevent this through hardware mechanisms. The host processor or accelerator must indicate that they have completed using the \emph{Sidebar} by writing to a hardware register before the other may proceed.

\subsection{Fine-Grained Cooperation}\label{ssec:coop}
In the same way that DMA is used at the beginning and end of accelerator tasks, the \emph{Sidebar} can be used to pass data between the accelerator and CPU \textit{during} an accelerator task. Combining this data-passing mechanism with a polling mechanism (detailed further in Section \ref{ssec:integration}) allows the accelerator to efficiently pass \textit{intermediate} results to the host processor and invoke desired functions on these results.

Fine-grained accelerator-CPU cooperation allows for improved performance and flexibility. Performance is improved as the accelerator will invoke the CPU for functions which are either not easily implemented in hardware (saving on area and power in the accelerator) or which run more slowly in hardware than on the CPU (such as non-linear activation functions). Flexibility is improved because difficult or costly hardware implementations of functions can be avoided in-lieu of performing the same function on the highly programmable host CPU. This paradigm of computing is not currently possible given the high overheads associated with accelerator data movement.

\subsection{Host System Integration}\label{ssec:integration}
In order for the host CPU to be able to collaborate with the accelerator, there must be a mechanism for the accelerator to call on the CPU to perform a piece of work. For our project, we plan on ignoring the complexities of integrating such a system into the context of a full operating system. Instead, we elect to have a simplified polling approach running on the host CPU as the sole application. The host will keep a table of functions the accelerator may call on the CPU to perform. These functions will be part of the accelerator's driver and will therefore be written and compiled ahead of time and reside in the host's memory.

When the accelerator wishes to invoke the CPU to perform a computation, the accelerator must first write the data needed for the computation in the \emph{Sidebar}. Once the data has been written, the accelerator will write the arguments of the computation to a specific set of \emph{Sidebar} locations. These arguments will include variables such as function pointers to host functions, pointers to data in the \emph{Sidebar}, and other information required for the invocation of the host. Once the data and arguments have been written into the \emph{Sidebar}, the accelerator writes to a specific \emph{Sidebar} location that the host is pulling on. This will signal to the host to begin the computation. The return process is similar to the invocation, except that the host will be setting up data and the accelerator will be waiting for the flag location to be pulled low.

\subsection{Coherence Interactions}\label{ssec:coherence}

When the accelerator is performing an acceleration task and invoking the host CPU, data must be placed into the \emph{Sidebar} by the accelerator before notifying the CPU of its task. The mechanism for this data movement is discussed in \ref{ssec:DMA}. The CPU will then operate on this data, potentially bringing it into its local registers. This data should not enter the cache hierarchy, however, since this intermediate result of accelerator computation is not normally application visible. Because this data should not enter the cache hierarchy and instead remains resident only in registers or within the \emph{Sidebar}, no coherence concerns are present.

Initial and final data movement to and from the accelerator's private memory are handled by DMA. This is the current protocol on many existing implementations of heterogeneous systems. 

\subsection{Consistency Interactions}\label{ssec:consistency}
In out-of-order host CPUs, depending on the consistency model, it may be possible for the status flag to be written before the return data has been written to the \emph{Sidebar}, even if the flag is written last in program order. To account for this, there are two possible solutions. The first is to have a separate load-store queue for \emph{Sidebar} memory instructions. This would allow for the system architects to decide on a consistency model specifically for the \emph{Sidebar}. However, additional fence instructions for the \emph{Sidebar} memory operations would then be required.

The other solution is to utilize existing load-store queues in the host processor. This means that the \emph{Sidebar} operations would obey the same consistency model as the host and would therefore be able to utilize existing fence instructions to maintain desired functionality. We see this as the optimal solution since it involves the least amount of modifications to the CPU microarchitecture.


\section{Design Overview}
\label{sec:design}

With Section \ref{sec:considerations} mentioning the base components and interactions of \textit{Sidebar}, it is useful to take a look at the bigger picture and see how \textit{Sidebar} fits into a real work flow in order to improve performance.

\textit{Sidebar} is best suited to workloads which are both a strong candidate for hardware acceleration but also contain "CPU-friendly" functionality - that which is \textit{better-suited} for execution on a powerful, general-purpose core. \textit{Sidebar} is also applicable to workloads which desire fine-grained cooperation between an accelerator and host, or those applications which desire a high level of flexibility for future algorithmic changes. Our work shows that neural network operations are a prime candidate for use with \textit{Sidebar}, but many other workloads would benefit from fine-grained cooperation.

Once a target algorithm is identified, an accelerator must be built for that task. This could be at the complexity level of a matrix multiplication kernel or could be a more abstract primitive like an entire convolution kernel. The accelerator is augmented with a finite state machine (FSM) and interface signals (data and control) capable of: (1) receiving commands from the host through a driver and (2) sending commands to the host in order to invoke CPU acceleration. In this work, gem5-Aladdin\cite{aladdin} is used to model this interaction  allowing us to combine accelerators with a CPU simulation infrastructure.

Once the accelerator hardware is completely built, a driver is created which allows for communication of data and tasks to the accelerator. Discussed in Section \ref{ssec:integration}, a \textit{Sidebar} implementation requires this driver both for starting and stopping the accelerator, but also for the accelerator to interface with the host CPU and invoke host operations through the \textit{Sidebar}.

In order to control the communication of intermediate data, \textit{Sidebar} dedicates a portion of the private, shared memory to accelerator-host communication. More specifically, the host CPU polls the driver-defined \textit{Sidebar} memory locations checking for flags which indicate a CPU task being invoked by the accelerator. When these flags are set, the CPU finds a function pointer in a dedicated memory location (as described in Section \ref{ssec:integration}) which tells the CPU which function it should perform on the contents of the shared memory region.

As the CPU is performing its computation, the accelerator FSM will be polling another region of the scratchpad waiting for the CPU to signal it has completed the work. While this communication is not ideal because it slightly reduces the usable scratchpad space and requires the host CPU to spin and wait, an interrupt-based mechanism might be used to solve both of these issues.

For more complex functions, a sea of multiple accelerators can be built. The obvious option is to build a monolithic accelerator that is designed to perform multiple activation functions. This configuration offers some flexibility, but wastes area and power on potentially unneeded hardware resources. One could build a set of accelerators which do not contain activation hardware, but these accelerators will need to pass intermediate results to the host processor through DMA incurring additional execution time and energy. Finally, could build an accelerator which does not contain activation hardware, but can communicate through a \textit{Sidebar}. This set of accelerators can invoke the host CPU to compute the activation functions of the network, passing data through the \textit{Sidebar} instead of DMA.

Through the use of these mechanisms and design flow, \textit{Sidebar} allows the host CPU to quickly respond to task requests from an accelerator attached to the system. This enables fine-grained cooperation between the CPU and potentially many accelerators with low overhead for communication and reduced energy consumption.
\section{Implementation}
\label{sec:impl}

\begin{figure*}[h]
\centering
\includegraphics[keepaspectratio,width=\linewidth]{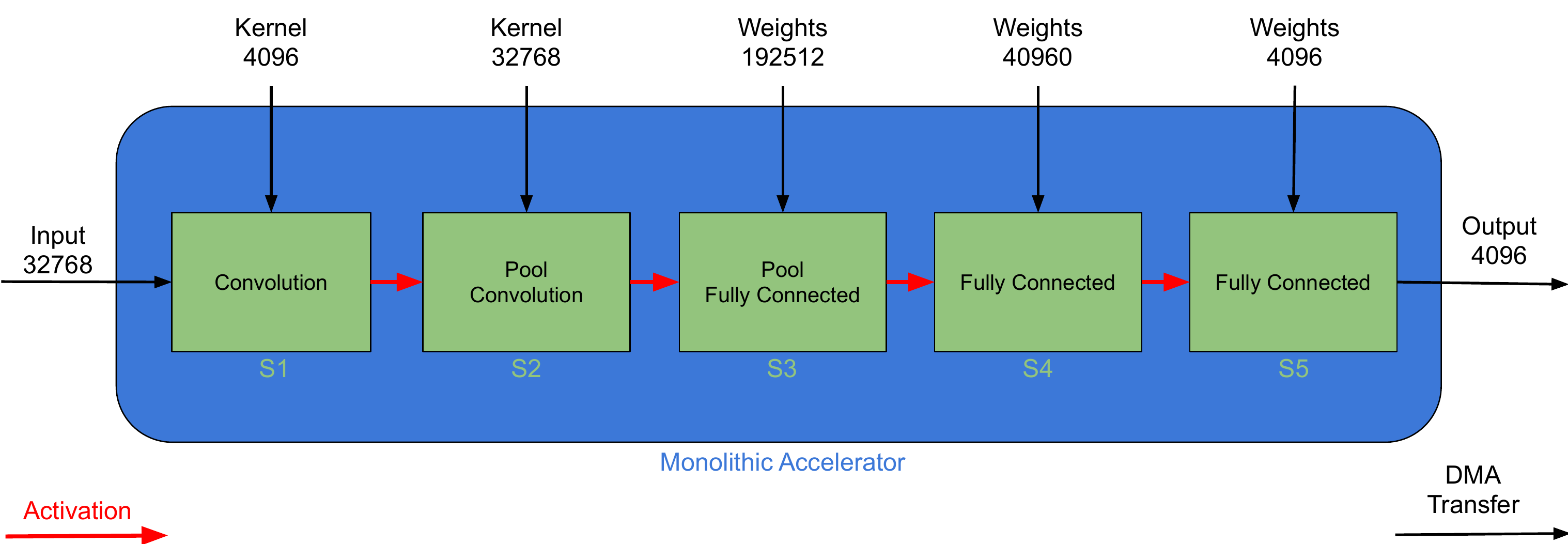}
\caption{\label{fig:accel}Lenet Accelerator Models}
\end{figure*}

The complete implementation and evaluation was done on gem-aladdin~\cite{aladdin}.
The source code is available at~\cite{gitlab}. Major components of the system are described further in this section.

\subsection{Gem5 System}
We use the default gem5 parameters from gem5-aladdin. The core parameters are defined in Table~\ref{tab:param_gem5}.

\begin{table}[htbp]
\caption{Gem5 Parameters}
\label{tab:param_gem5}
\begin{center}
\begin{tabular}{|l|l|}
\hline
Component  & Gem5 Parameter        \\ \hline
CPU        & Single Core DerivO3CPU \\ \hline
Memory     & 4GB DDR3\_1600\_8x8 \\ \hline
Clock      & 1 GHz \\ \hline
\end{tabular}%
\end{center}
\end{table}

\emph{Limitations:} gem5-aladdin only supports system call emulation mode for program execution. In this mode programs are executed without an OS layer. Any systems calls are functionally emulated by gem5. Due to this limitation we do not implement the OS dependent interrupt interface. The program flow is completely controlled by the application running on main processor.

\subsection{Accelerators}
The basis for our simulations was a neural network model in the Lenet style~\cite{lenet,alexnet}. The exact model was adapted from one in the Pytorch documentation~\cite{lenet_source} specifically developed to classify CIFAR-10~\cite{cifar}. Some of the hyper-parameters were modified for simulation purposes. It consists of two convolutional layers, each followed by an activation and a pooling layer. These are then followed by three fully connected layers, with activations in-between. The complete network was implemented in two distinct forms as shown in Figure~\ref{fig:accel}.

\begin{table}[htbp]
\caption{Accelerator Parameters}
\label{tab:param_accel}
\small
\begin{tabular}{|l|l|l|l|}
\hline
Accelerator         & Cycles                       & Energy (Cycles$\times$mW) & Area (uM$^2$) \\ \hline
Relu Monolithic     & \multicolumn{1}{|r|}{122151} & \multicolumn{1}{|r|}{724294354}        & 4.82445e+08 \\ \hline
SoftPlus Monolithic & \multicolumn{1}{|r|}{147967} & \multicolumn{1}{|r|}{873817638}        & 4.82448e+08 \\ \hline
S1                  &  \multicolumn{1}{|r|}{23124} & \multicolumn{1}{|r|}{138988189}        & 4.61686e+08 \\ \hline
S2                  &  \multicolumn{1}{|r|}{22541} &  \multicolumn{1}{|r|}{86039447}        & 2.90202e+08 \\ \hline
S3                  &  \multicolumn{1}{|r|}{66060} &  \multicolumn{1}{|r|}{51164791}        & 6.10141e+07 \\ \hline
S4                  &  \multicolumn{1}{|r|}{17847} &   \multicolumn{1}{|r|}{3560833}        & 1.46956e+07 \\ \hline
S5                  &   \multicolumn{1}{|r|}{2546} &    \multicolumn{1}{|r|}{110980}        & 2.60089e+06 \\ \hline

\end{tabular}%
\end{table}

\subsubsection{Monolithic}
The monolithic version implements the complete network in a single accelerator. Consider the blue box in Figure~\ref{fig:accel}. All layers and activations are within this monolithic accelerator. The black arrows represent data motion between the accelerator and main memory. All data transfers are DMA. We use different activation functions in this accelerator. A comparison between Relu and SoftPlus is shown in Table~\ref{tab:param_accel}. These two activation functions were chosen because Relu is the most commonly used and SoftPlus is the most computationally complex.

\subsubsection{Small Primitives}
Consider Figure~\ref{fig:accel} again. For this configuration we define layers without intervening activations as small accelerator primitives. These are represented as green boxes in Figure~\ref{fig:accel}. In this configuration, the activations are computed on the main processor. Hence the activations convert to data transfer back to processor memory and a computation on the processor. The data transfer may be via DMA or a low latency \emph{sidebar} based transfer. The realized parameters for the small accelerators are also shown in Table~\ref{tab:param_accel}

\emph{Limitation:} Gem5 integrates with Aladdin via an ioctl interface. Given this limited interface we could not fully implement low latency cross communication between the two simulation domains. Hence we approximate the evaluations by synthetically controlling the latency of data transfers between Gem5 and Aladdin simulation domains for the activation calls only. Specifically, the access on accelerator side are from its local scratchpad. CPU side accesses are large contiguous memory operations, which, with prefetching reach cache level latency. Hence these accesses emulate indirectly the latency to access the \emph{sidebar}. Note that the input, output and parameters still use DMA based data transfers. \emph{Sidebar} like latencies are used only to provide intermediate data to host processor to compute activations and by the next accelerator to access the results of processor based activation computations.
The network itself is untrained and hence does not provide meaningful data outputs. Hence the fact that Aladdin and Gem5 incurr Sidebar latency costs without the simulations actually communicating data does not impact the accuracy of our results.

\subsection{Scenarios}

\begin{figure*}[h]
\centering
\includegraphics[keepaspectratio,width=\linewidth]{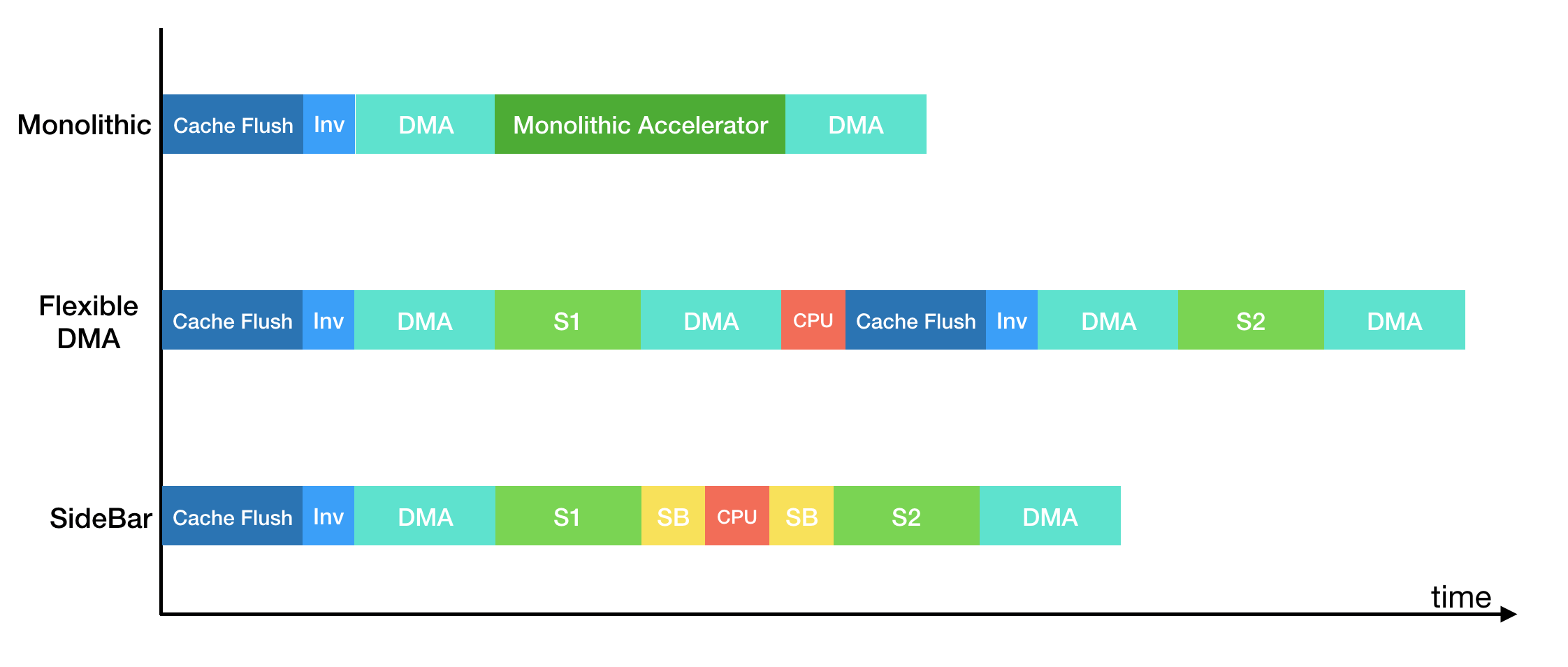}
\caption{\label{fig:timeline}Execution Timeline}
\end{figure*}

Based on the implementations described above we find three particular scenarios of interest which form the basis of our evaluation, as shown in Figure~\ref{fig:timeline}.

\subsubsection{Monolithic}
The monolithic accelerator implements all layers of the neural network including activation functions in a single accelerator. The execution begins by the host CPU flushing its caches to DRAM and then invalidating the cache lines. Now the actual DMA can begin. The CPU initiates the DMA load and the accelerator receives the data. The accelerator then performs its computation. Once complete, the accelerator then performs a DMA store. The CPU can then read the accelerator result and use it accordingly.

\subsubsection{Flexible DMA}
For flexible DMA, we wanted to understand how an SoC would need to leverage DMA to achieve the same level of flexibility as \emph{Sidebar}. For this accelerator structure, the neural network accelerator is split into five different accelerators, corresponding to each of the network layers, excluding activation functions. The initial and final DMA transfer processes are the same as the monolithic accelerator. However, since each accelerator is separate, this process must be replicated for each invocation of an accelerator. The benefit of such a system is that the activation functions are performed on the CPU between DMAs. This allows the network activation functions to be changed very easily and also implemented in software. Existing machine learning accelerators implement a subset of activation functions in hardware and have no mechanism to introduce new functions. The downside of this programmability is that the communication overhead is rather high. Something that our results do not show however, is that breaking up the accelerators would allow for pipelineing of computations. It would be important to note that all of the accelerators attempting to DMA to and from the CPU simultaneously would most likely demonstrate a communication bottleneck.

\subsubsection{Sidebar}
The goal of Sidebar is to accomplish the same level of programmability as the flexible DMA accelerators with reduced communication overhead. The reduction in communication costs comes from the use of Sidebars between the accelerators and the host CPU. Using Sidebars allows us to forgo the cache flushing and invalidation costs of using DMA. Using Sidebar also allows for faster data transfers since Sidebar sits at the L1 level in the memory hierarchy. This means that the CPU and accelerators need not go to DRAM to get data for DMA. For this work, Sidebar is use to eliminate the intermediate DMA transactions between host and accelerator. The initial and final DMA processes must still take place.
\section{Evaluation}
\label{sec:eval}

The flexibility and performance of \emph{Sidebar} are evaluated using a neural network inference pass as a workload. In this workload, a host CPU sets up the acceleration task(s) by allocating memory and mapping various arrays. The host then invokes the accelerator(s) and waits until the task is complete before checking the output for correctness. With multiple network layers and activation functions between them, this workload perfectly fits the model of fine-grained cooperation between accelerators and host processors.

Each time the workload is run, Gem5 collects statistics for the inference's execution. These statistics include performance and power numbers for the accelerators in the system, as well as information about the system interconnect and its traffic. Using these statistics, we evaluate the performance, communication, and energy of \emph{Sidebar}.

\subsection{Latency}

\begin{figure}[ht]
\centering
\includegraphics[width=\linewidth]{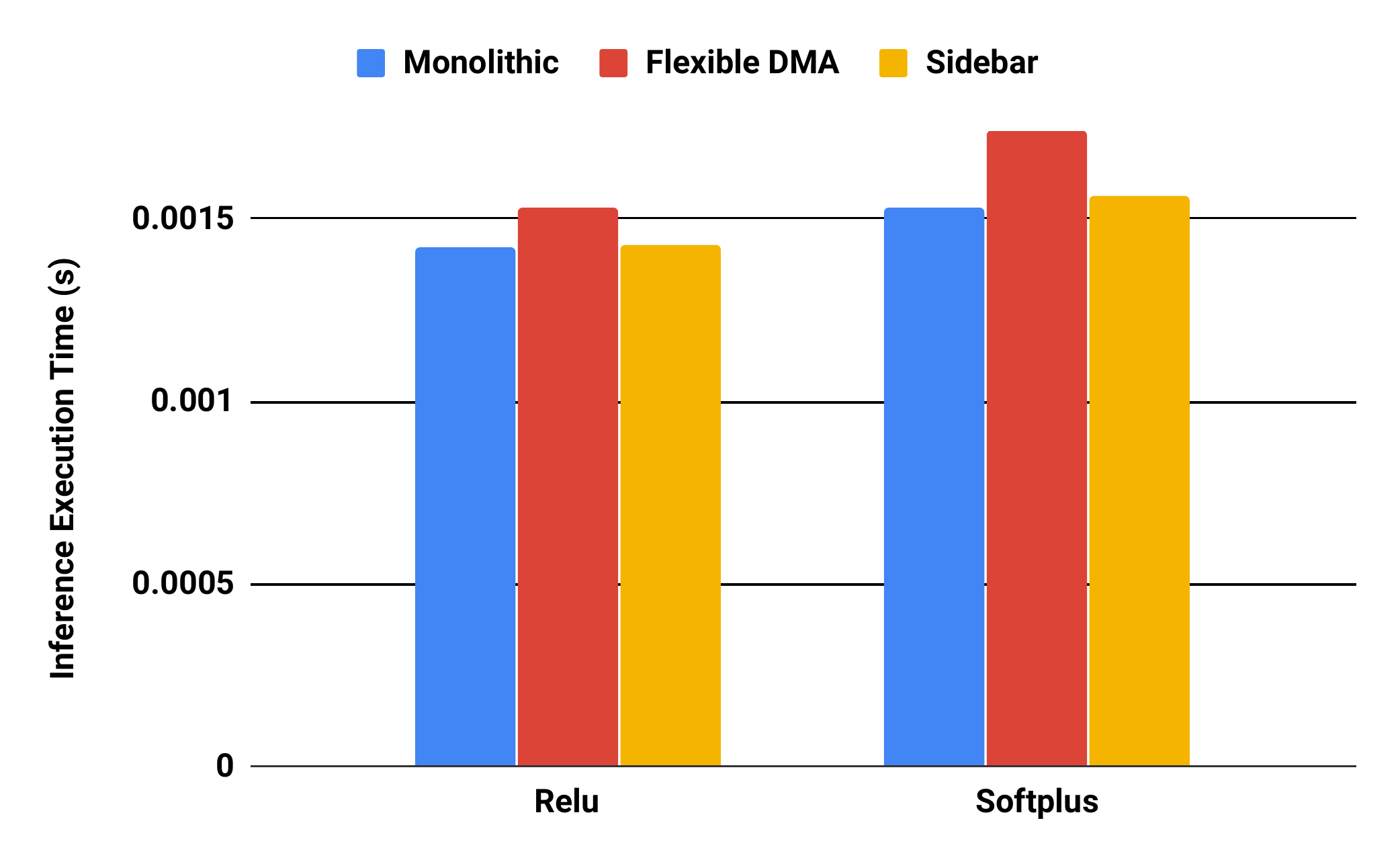}
\caption{Inference latency of Lenet convolutional neural network with hardware acceleration enabled. Monolithic refers to a single, inflexible accelerator. Flexible DMA refers to a flexible set of accelerators with DMA only for communication. \emph{Sidebar} represents our implementation in Gem5 + Aladdin.}
\label{fig:perf}
\end{figure}

Shown in Figure \ref{fig:perf} are the performance results of the two baseline designs and our \emph{Sidebar} implementation. Shown is the latency of a single inference pass.

From the figure, we can see that increased flexibility comes at a cost. Both the flexible DMA baseline and \emph{Sidebar} incur slight overheads during execution. The flexible DMA configuration has a run time which is 8 to 14 percent longer than the monolithic accelerator, while \emph{Sidebar} manages to stay within 2 percent of the monolithic accelerator's performance.

Furthermore, the testing of two different activation functions shows that offloading complex computation to the host CPU is a viable alternative to an expensive hardware implementation as in the monolithic accelerator. We can see that for the more complex activation, softplus, \emph{Sidebar} allowed for better cooperation with lower overhead than DMA. This is evidenced by the widening delta between the flexible DMA configurations while the \emph{Sidebar} desin shows consistent performance relative to the monolithic design.

\subsection{Data Communication}

After assessing the performance of \emph{Sidebar}, we turned to the evaluation of each system's energy consumption. This evaluation was performed using data from CACTI \cite{cacti:muralimanohar2009cacti} as well as statistics on data transferred within each system. There are two routes for data transfer. The first is the system or DRAM bus which is where all DMA transfers take place. The second route is the \emph{Sidebar} implementation which we model as a tightly coupled storage array connected between the host CPU and accelerator pool.

\begin{figure}[ht]
\centering
\includegraphics[width=\linewidth]{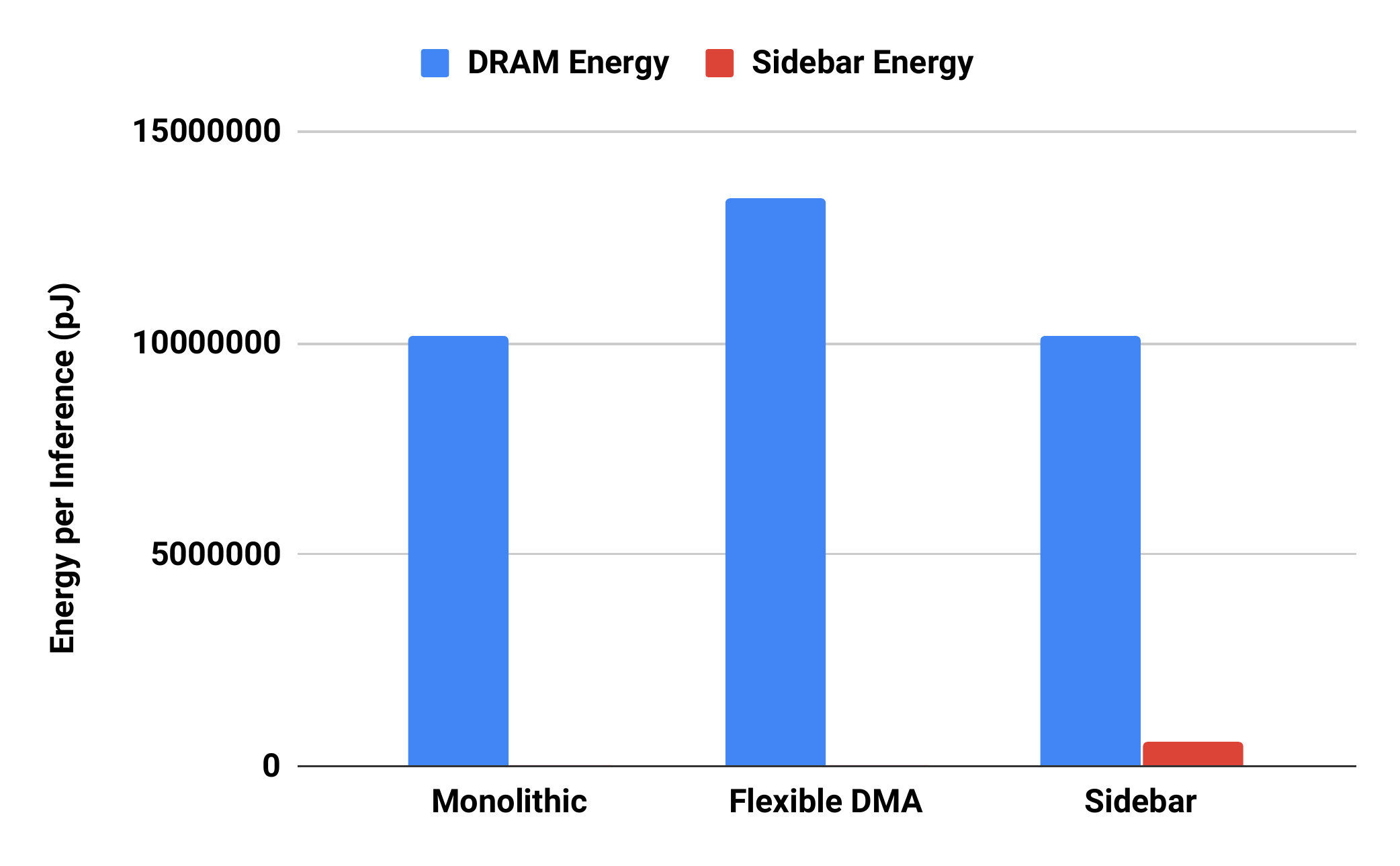}
\caption{Data communication energy in the tested accelerator configurations. DRAM energy refers to data moved on the system bus while \emph{Sidebar} energy refers to data moved through the \emph{Sidebar}.}
\label{fig:bus}
\end{figure}

The results shown in Figure \ref{fig:bus} depict the merits of specialization. The use of simple yet generic DMA operations for data transfer leads to a huge amount of data being sent on the system bus in the flexible DMA configuration. This leads to the flexible DMA design using 32 percent more energy per inference than the monolithic accelerator which can keep inter-layer data transfers internal to its data path for improved efficiency. \emph{Sidebar} incurs only 6 percent more energy consumption from data movement than the monolithic design. This is because \emph{Sidebar} alleviates some of the energy consumption incurred by moving data between the accelerator and host CPU by transferring the data through the private memory store. This dramatically reduces dynamic energy and allows the \emph{Sidebar} configuration to offer the flexibility of DMA transfers with nearly the energy consumption of a monolithic implementation.

\subsection{Normalized Energy}

After looking at system performance and system energy, a useful final metric to consider is energy-delay product (EDP) which allows one to compare the energy efficiency of designs with varying performance and power consumption statistics. EDP is the product of the computation run time with the energy consumed during the execution of the workload. Smaller values are better and signify that a design is very high performance, very power efficient, or a strong balance of the two.

\begin{figure}[ht]
\centering
\includegraphics[width=\linewidth]{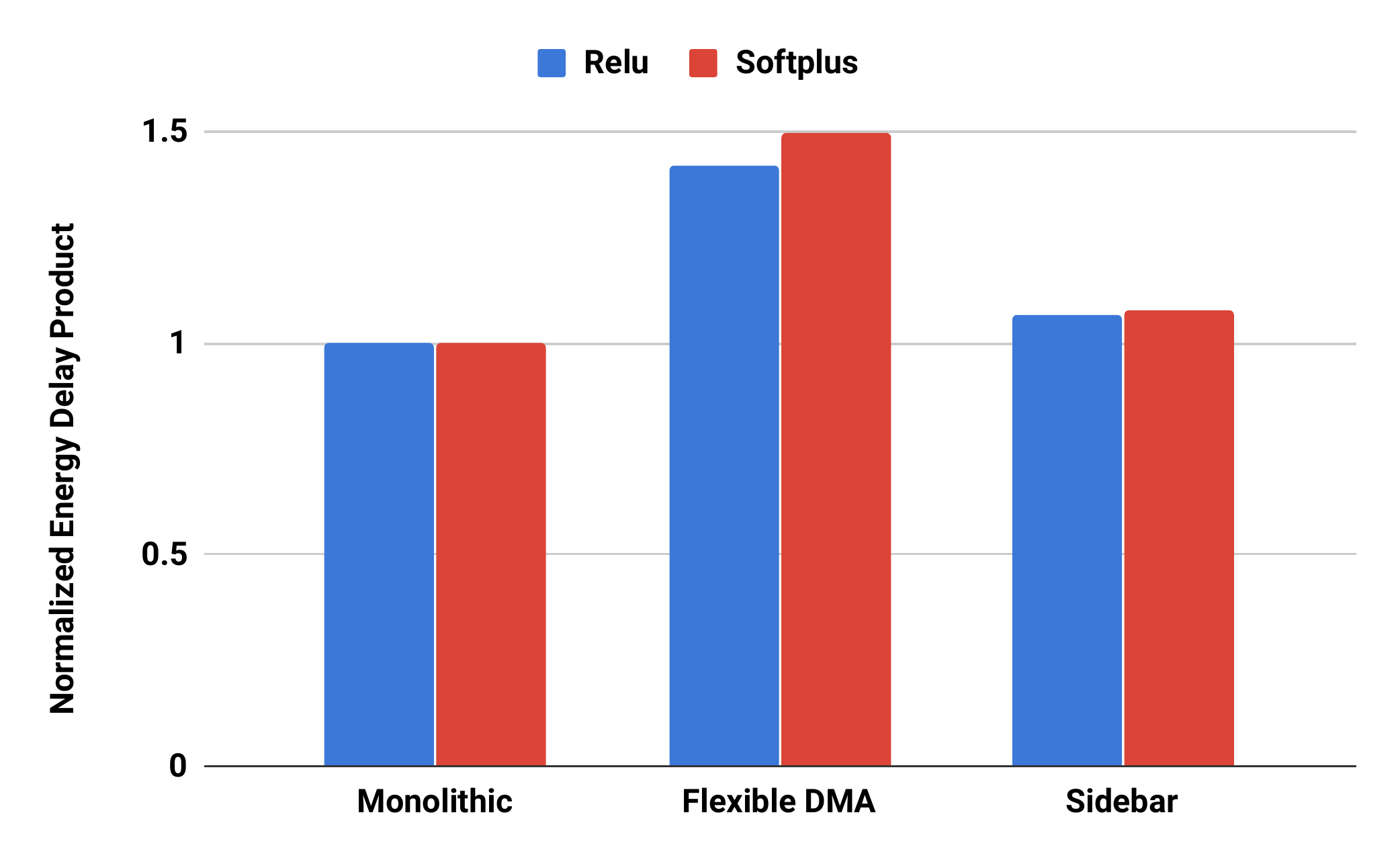}
\caption{Normalized energy consumption}
\label{fig:norm}
\end{figure}

Figure \ref{fig:norm} presents the EDP of each design normalized to the monolithic accelerator. Due to the combination of sizable DMA transfer overheads and the increased overall data movement of the flexible DMA design, a nearly 50 percent increase is EDP can be seen compared with the monolithic design. \emph{Sidebar}, on the other hand, has only a slight increase in EDP when compared with the monolithic accelerator. This stems from the fact that \emph{Sidebar} has comparable performance \textit{and} a greatly reduced amount of high-energy bus communication relative to the flexible DMA design. This means \emph{Sidebar} sees only a 7 percent increase in EDP compared to the monolithic design, and is nearly 40 percent better than the flexible DMA configuration.

\subsection{Discussion}

The overall performance and energy evaluations performed with \emph{Sidebar} are quite encouraging. Through the use of a tightly-coupled storage mechanism, \emph{Sidebar} enables low-cost cooperation between a host CPU and fixed-function hardware accelerators. Our experiments have shown that \emph{Sidebar} offers the flexibility of a DMA-based system with performance and energy consumption competitive with monolithic accelerator designs. \emph{Sidebar}'s private scratchpad offers dramatically reduced energy when compared to the high-capacitance system memory bus. Because of the tight coupling between the host and accelerator, \emph{Sidebar} also offers fast access to data for cooperative tasks.

Indeed, this level of performance and energy efficiency for cooperative workloads has not been shown before. \emph{Sidebar} provides a glimpse at the future of cooperative workload execution where a host CPU dictates tasks to a whole pool of accelerators. \emph{Sidebar} offers great specialization with its support for hardware accelerators, but provides programmers with the flexibility to update and modernize their code as new libraries and activation functions are developed while also enabling a new paradigm of low-cost accelerator cooperation.
\section{Future Work}
While we have shown that there are compelling gains in terms of system flexibility for our solution, this is only the beginning of the potential for \emph{Sidebar}. The first area we consider for future work is using \emph{Sidebar} to stream working data to/from accelerators. This could theoretically decrease the latencies of the initial and final DMAs. However, it would require potentially much larger \emph{Sidebar}s, or much smarter communication methods. Future work could reasonably invent a method for streaming data through \emph{Sidebar} to initialize accelerator storage.

\begin{figure}[ht]
\centering
\includegraphics[width=\linewidth]{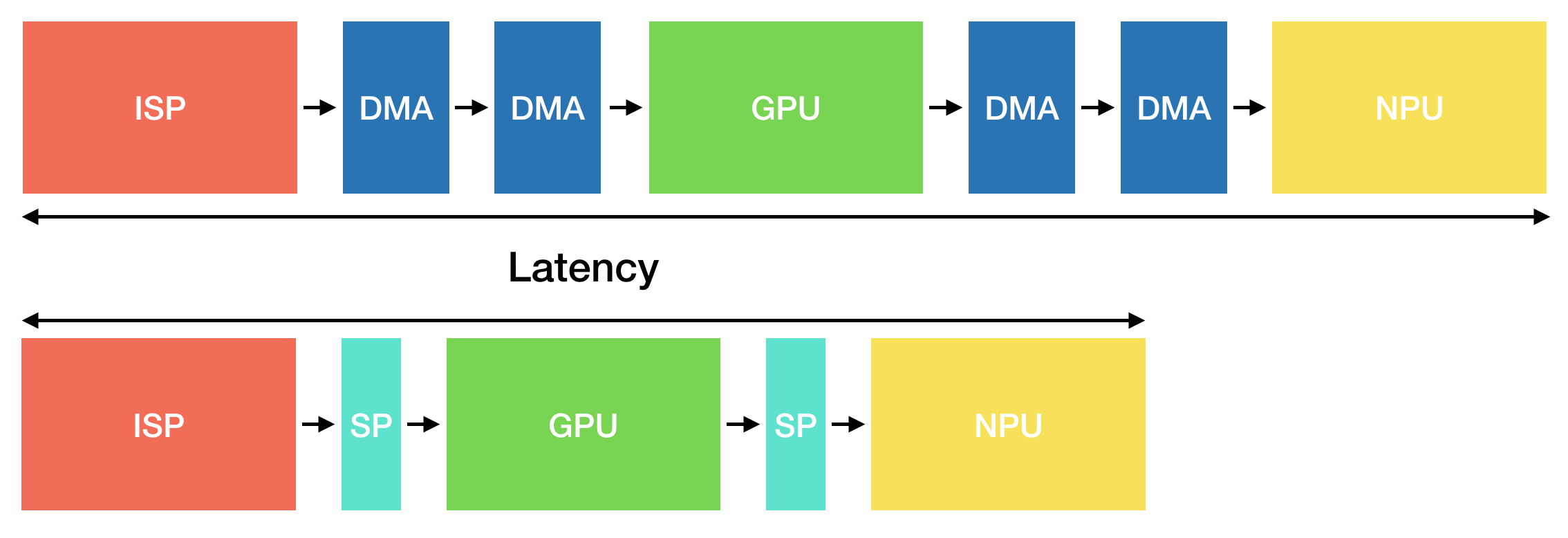}
\caption{Latency of image processing, GPU pre-processing and neural network inference. With intermediate CPU DMAs (top), with intermediate Sidebar usage (bottom)}
\label{fig:latency}
\end{figure}

The main idea of this work is that we are reducing the cost of communication between accelerator and host. This work could rather easily be expanded to incorporate accelerator-to-accelerator communication as well. This could be potentially very useful in collaborative workloads between accelerators without requiring the intervention of the host processor. Consider a modern SoC with an image processing pipeline and some sort of machine learning acceleration. The image may first begin in a demosaicing accelerator (converting RAW sensor data to pixel data), but then may be merged with other images in an HDR processing accelerator. Finally, there may be some additional processing that requires the GPU before the image is sent to the neural network accelerator. With modern memory systems, these accelerators must communicate through the CPU using DMA to stream data back and forth. While systems can pipeline such data transfers to amortize the cost of the transfer delays, such processing pipelines are frequently common in self-driving car hardware, where latency is somewhat of a priority. Thus, being able to communicate through an accelerator-to-accelerator \emph{Sidebar} without the overhead of flushes, invalidations, and higher levels of the memory hierarchy would prove very beneficial in terms of latency.

Since the communication costs are theoretically very small, \emph{Sidebar} could promote a sea of primitive accelerators. Rather than having few accelerators that accelerate large tasks, an SoC can implement many accelerators that implement small, reusable tasks. For example, it may be possible to create a general convolution accelerator that can be reused between image processing and machine learning pipelines. Normally, the overhead of invoking many small accelerators and the associated passing of data would be too significant. However, with the work shown here, this type of work may be feasible since these overheads are significantly reduced. The only potential drawback of separating accelerators as such would require each accelerator to have its own local memories. We noticed this storage duplication in our evaluation. When separating out each neural network layer into its own accelerator, the area grew significantly because each accelerator required its own private memory. This could be mitigated by allowing some sort of private memory sharing between accelerators. It could potentially even be possible to reuse \emph{Sidebar} as an accelerator scratchpad, allowing for a rather area efficient solution.
\section{Related Work}

A major problem with hardware accelerator development is the lack of standardized interfaces. Previous works have focused on identifying this problem and attempt to solve it in various ways~\cite{interfaceexploration,spandex, capi}.

In this project we define a standard interface to hardware accelerators that is independent of the nature of the accelerator.
This interface uses a scratchpad memory between the processor and hardware accelerator.
ScratchPad memory has been well studied in literature with works proposing basic~\cite{scratchpadmem} to advanced~\cite{stash} usage of the scratchpad.

While a cache could also be placed between the processor and hardware accelerator, caches incur high overheads from two mechanisms. First, caches require additional hardware to store and compare tags for the data held within. Second, caches require address translation when the host processor and accelerator are operating in different domains. We believe these overheads are unnecessary since we intend to focus on fixed-function hardware accelerators. Since these accelerators rely upon a known data layout in memory, the host and accelerator can avoid checking tags by ensuring data is placed into and read from the correct locations. Further, the data we intend to pass between the host processor and accelerator is not normally application visible and therefore has no need to enter the application memory space. Using physical scratchpad addresses therefore offers a low-overhead approach to fine-grained data movement and cooperation.

Flexible accelerator design has also been addressed by many previous works. \cite{npu1,npu2}~provide a programmable layer structure to emulate a variety of networks. But the more recent notion is to decompose networks to some common primitives which can be then freely used to accelerate different network models. \emph{Sidebar} technique fits well with this kind of design principle. But we go further and include the host processor into the neural network acceleration hardware.

\section{Conclusion}
\label{sec:conclusion}
In this paper we explore flexible accelerator design for neural networks. The design uses static tensor computation units with a low latency \emph{Sidebar} to allow some computations to be offloaded to host processor. Compared to large fixed function accelerators our design achieves similar performance while being flexible in design. We believe such designs will be required to develop accelerators for applications that undergo rapid development but still require the performance and energy improvements of hardware acceleration.

\bibliographystyle{ieeetr}
\bibliography{ref}

\begin{thebibliography}{10}

\bibitem{hwaccel}
M.~{Porrmann}, U.~{Witkowski}, H.~{Kalte}, and U.~{Ruckert}, ``Implementation
  of artificial neural networks on a reconfigurable hardware accelerator,'' in
  {\em Proceedings 10th Euromicro Workshop on Parallel, Distributed and
  Network-based Processing}, pp.~243--250, Jan 2002.

\bibitem{tensorflow}
M.~Abadi, P.~Barham, J.~Chen, Z.~Chen, A.~Davis, J.~Dean, M.~Devin,
  S.~Ghemawat, G.~Irving, M.~Isard, {\em et~al.}, ``Tensorflow: A system for
  large-scale machine learning,'' in {\em 12th $\{$USENIX$\}$ Symposium on
  Operating Systems Design and Implementation ($\{$OSDI$\}$ 16)}, pp.~265--283,
  2016.

\bibitem{tpu}
N.~P. {Jouppi}, C.~{Young}, N.~{Patil}, D.~{Patterson}, G.~{Agrawal},
  R.~{Bajwa}, S.~{Bates}, S.~{Bhatia}, N.~{Boden}, A.~{Borchers}, R.~{Boyle},
  P.~{Cantin}, C.~{Chao}, C.~{Clark}, J.~{Coriell}, M.~{Daley}, M.~{Dau},
  J.~{Dean}, B.~{Gelb}, T.~V. {Ghaemmaghami}, R.~{Gottipati}, W.~{Gulland},
  R.~{Hagmann}, C.~R. {Ho}, D.~{Hogberg}, J.~{Hu}, R.~{Hundt}, D.~{Hurt},
  J.~{Ibarz}, A.~{Jaffey}, A.~{Jaworski}, A.~{Kaplan}, H.~{Khaitan},
  D.~{Killebrew}, A.~{Koch}, N.~{Kumar}, S.~{Lacy}, J.~{Laudon}, J.~{Law},
  D.~{Le}, C.~{Leary}, Z.~{Liu}, K.~{Lucke}, A.~{Lundin}, G.~{MacKean},
  A.~{Maggiore}, M.~{Mahony}, K.~{Miller}, R.~{Nagarajan}, R.~{Narayanaswami},
  R.~{Ni}, K.~{Nix}, T.~{Norrie}, M.~{Omernick}, N.~{Penukonda}, A.~{Phelps},
  J.~{Ross}, M.~{Ross}, A.~{Salek}, E.~{Samadiani}, C.~{Severn}, G.~{Sizikov},
  M.~{Snelham}, J.~{Souter}, D.~{Steinberg}, A.~{Swing}, M.~{Tan},
  G.~{Thorson}, B.~{Tian}, H.~{Toma}, E.~{Tuttle}, V.~{Vasudevan}, R.~{Walter},
  W.~{Wang}, E.~{Wilcox}, and D.~H. {Yoon}, ``In-datacenter performance
  analysis of a tensor processing unit,'' in {\em 2017 ACM/IEEE 44th Annual
  International Symposium on Computer Architecture (ISCA)}, pp.~1--12, June
  2017.

\bibitem{Cybenko1989}
G.~Cybenko, ``Approximation by superpositions of a sigmoidal function,'' {\em
  Mathematics of Control, Signals and Systems}, vol.~2, pp.~303--314, Dec 1989.

\bibitem{telgarsky16}
M.~Telgarsky, ``Benefits of depth in neural networks,'' {\em CoRR},
  vol.~abs/1602.04485, 2016.

\bibitem{alexnet}
A.~Krizhevsky, I.~Sutskever, and G.~E. Hinton, ``Imagenet classification with
  deep convolutional neural networks,'' in {\em Advances in Neural Information
  Processing Systems 25} (F.~Pereira, C.~J.~C. Burges, L.~Bottou, and K.~Q.
  Weinberger, eds.), pp.~1097--1105, Curran Associates, Inc., 2012.

\bibitem{interfaceexploration}
P.~{Possa}, D.~{Schaillie}, and C.~{Valderrama}, ``Fpga-based hardware
  acceleration: A cpu/accelerator interface exploration,'' in {\em 2011 18th
  IEEE International Conference on Electronics, Circuits, and Systems},
  pp.~374--377, Dec 2011.

\bibitem{spandex}
J.~Alsop, M.~D. Sinclair, and S.~V. Adve, ``Spandex: A flexible interface for
  efficient heterogeneous coherence,'' in {\em Proceedings of the 45th Annual
  International Symposium on Computer Architecture}, ISCA '18, (Piscataway, NJ,
  USA), pp.~261--274, IEEE Press, 2018.

\bibitem{shao2015toward}
Y.~S. Shao, S.~Xi, V.~Srinivasan, G.-Y. Wei, and D.~Brooks, ``Toward
  cache-friendly hardware accelerators,'' in {\em HPCA Sensors and Cloud
  Architectures Workshop (SCAW)}, pp.~1--6, 2015.

\bibitem{aladdin}
Y.~S. {Shao}, S.~{Xi}, V.~{Srinivasan}, G.~Y. {Wei}, and D.~{Brooks},
  ``Co-designing accelerators and soc interfaces using gem5-aladdin,'' {\em
  IEEE/ACM International Symposium on Microarchitecture}, vol.~49, Oct 2016.

\bibitem{gitlab}
E.~L. B.~S. A.~Bansal, C.~Coats, ``{Project source code}.''
  \url{https://gitlab.engr.illinois.edu/ayooshb2/gem5-aladdin}, 2019.

\bibitem{lenet}
Y.~{Lecun}, L.~{Bottou}, Y.~{Bengio}, and P.~{Haffner}, ``Gradient-based
  learning applied to document recognition,'' {\em Proceedings of the IEEE},
  vol.~86, pp.~2278--2324, Nov 1998.

\bibitem{lenet_source}
P.~Tutorials, ``{Cifar10 tutorial}.''
  \url{https://github.com/pytorch/tutorials/blob/master/beginner_source/blitz/cifar10_tutorial.py},
  2019.

\bibitem{cifar}
A.~Krizhevsky and G.~Hinton, ``Learning multiple layers of features from tiny
  images,'' tech. rep., Citeseer, 2009.

\bibitem{cacti:muralimanohar2009cacti}
N.~Muralimanohar, R.~Balasubramonian, and N.~P. Jouppi, ``Cacti 6.0: A tool to
  model large caches,'' {\em HP laboratories}, pp.~22--31, 2009.

\bibitem{capi}
J.~{Stuecheli}, B.~{Blaner}, C.~R. {Johns}, and M.~S. {Siegel}, ``Capi: A
  coherent accelerator processor interface,'' {\em IBM Journal of Research and
  Development}, vol.~59, pp.~7:1--7:7, Jan 2015.

\bibitem{scratchpadmem}
R.~{Banakar}, S.~{Steinke}, , M.~{Balakrishnan}, and P.~{Marwedel},
  ``Scratchpad memory: a design alternative for cache on-chip memory in
  embedded systems,'' in {\em Proceedings of the Tenth International Symposium
  on Hardware/Software Codesign. CODES 2002 (IEEE Cat. No.02TH8627)},
  pp.~73--78, May 2002.

\bibitem{stash}
R.~{Komuravelli}, M.~D. {Sinclair}, J.~{Alsop}, M.~{Huzaifa}, M.~{Kotsifakou},
  P.~{Srivastava}, S.~V. {Adve}, and V.~S. {Adve}, ``Stash: Have your
  scratchpad and cache it too,'' in {\em 2015 ACM/IEEE 42nd Annual
  International Symposium on Computer Architecture (ISCA)}, pp.~707--719, June
  2015.

\bibitem{npu1}
H.~Esmaeilzadeh, A.~Sampson, L.~Ceze, and D.~Burger, ``Neural acceleration for
  general-purpose approximate programs,'' in {\em Proceedings of the 2012 45th
  Annual IEEE/ACM International Symposium on Microarchitecture}, pp.~449--460,
  IEEE Computer Society, 2012.

\bibitem{npu2}
T.~Moreau, M.~Wyse, J.~Nelson, A.~Sampson, H.~Esmaeilzadeh, L.~Ceze, and
  M.~Oskin, ``Snnap: Approximate computing on programmable socs via neural
  acceleration,'' in {\em 2015 IEEE 21st International Symposium on High
  Performance Computer Architecture (HPCA)}, pp.~603--614, IEEE, 2015.

\end{thebibliography}

\end{document}